\documentclass[twocolumn,showpacs,showkeys, superscriptaddress,aps]{revtex4-2}
\usepackage{latexsym}
\usepackage{amssymb}
\usepackage{graphicx}
\usepackage[colorlinks,linkcolor=blue,urlcolor=blue, anchorcolor=blue,citecolor=blue]{hyperref}
\usepackage{natbib}
\newcommand{\bq}{\begin{equation}}
\newcommand{\eq}{\end{equation}}
\newcommand{\bn}{\begin{eqnarray}}
\newcommand{\en}{\end{eqnarray}}
\begin{document}

\title{Singlet-doublet transitions and Josephson currents in a superconducting ring  with  a quantum dot }

\author { Guo-Hui Ding}
\affiliation{Key Laboratory of Artificial Structures and Quantum Control (Ministry of Education), School of Physics and Astronomy, Shanghai Jiao Tong University, 800 Dongchuan Road, Shanghai 200240, China}
\author {Fei Ye}
\affiliation{ Department of Physics, Southern University  of Science  and Technology, Shenzhen 518055, China}
\author {Bing Dong }
\affiliation{Key Laboratory of Artificial Structures and Quantum Control (Ministry of Education), School of Physics and Astronomy, Shanghai Jiao Tong University, 800 Dongchuan Road, Shanghai 200240, China}

\date{\today }

\begin{abstract}
We investigate the ground state properties of a superconducting ring embedded with a quantum dot (QD) by using a variational wave-function approach. A theoretical formulation for the treatment of the finite-U Anderson impurity  coupled with a superconducting ring are presented.  We demonstrate  singlet-doublet transitions of the ground state for this system with the QD in the mixed valence regime. It is shown that the supercurrent in the superconductor ring shows oscillations with the external enclosed magnetic flux and exhibits abrupt jumps at the singlet-doublet phase transition points.

\end{abstract}
\maketitle
\newpage

\section{introduction}
 In recent years, the electron transport properties of quantum dots (QDs) coupled with superconductors have attract a great deal of  research interests.  Many interesting phenomena in these systems have been observed or  predicted, e.g., the effects of Andreev bound state (ABS) \cite{Bordin2025}and multiple Andreev reflection (MAR) on electron transport at finite bias voltage;\cite{Buitelaar2002, *Buitelaar2003}  The quantum phase transition between $0$ and $\pi$ junctions (the singlet-doublet transition)\cite{Buitelaar2002, Rozhkov1999,Rozhkov2000, Choi2004, Kim2013, Li2017} driven by the competition between the superconducting  pairing and the Kondo correlation,  which leads to abrupt changes in Josepshon currents tunneling through the junctions.\cite{Eichler2009}  Recently, the singlet-doublet transition has also been detected  in the system with embedding a gate-controlled QD in the InAs/Al junction by using transmon circuit technique.\cite{Bargerbos2022,Bargerbos2023}  The existence and possible experimental detection of intermediate states ($0'$ and $\pi'$ phases) in the QD Josephson junctions was proposed theoretically.\cite{Rozhkov1999,Rozhkov2000,Lee2022}   It is expected that the ABS in the junction occupied by a quasiparticle can serve as a spin quibit, and coherent manipulation of this Andreev spin qubit and the spin dependent Josephson current will have potential applications in quantum information processing based on electron spins.\cite{Hays2021, Bargerbos2023}

The $0$-$\pi$ transition and Josephson currents in the QD Josephson junctions have been investigated theoretically based on a single impurity Anderson model with superconducting electrodes by using a variety of theoretical methods,\cite{Meden2019} including the perturbative theory in the weak tunnelling limit,\cite{Glazman1989,Spivak1991} and the nonpertubative methods, e.g., the Hartree-Fock approximation,\cite{Rozhkov1999} the variational wave function calculation,\cite{Rozhkov2000} the noncrossing approximation (NCA), \cite{Clerk2000}   the numerical renormalization group (NRG)\cite{Yoshioka2000, Oguri2004, Choi2004, Lee2022}, the functional RG\cite{Karrasch2008,Meden2019} and the continuous-time quantum Monte Carlo method,\cite{Luitz2010,Luitz2012} etc. In this paper, we will limit our attention to the variational wave function approach related to this problem. The variational wave method was initially applied to study the Kondo effect of the electrons in a conduction band scattered by a localized magnetic impurity spin (within the s-d model), and it was shown that this method does capture the essential nonperturbative physical properties of the Kondo effect. \cite{Yosida1966, Appelbaum1967}
Later, the variational  method was generalized to study the Anderson impurity model, \cite{Varma} the  mixed valence and  multi-orbital impurity  problem \cite{Gunnarsson1983}, and also the problem of a magnetic impurity in superconductors. \cite{Soda1967} Rozhkov and  Arovas \cite{Rozhkov2000} applied the wave function method to the QD Josephson junctions,  and obtained a phase diagram for the $0$-$\pi$ transition within the $U=\infty$  Anderson impurity model with superconducting electrodes. Recently, Ili\v{c}in and \v{Z}itko \cite{Ilicin2025} generalized the variational wave function approach to the finite-$U$ Anderson impurity model with superconducting electrodes in the particle-hole symmetric region, and investigated the crossover from ABS to Yu-Shiba-Rusinov state (YSR).

In this work, we apply the finite-$U$ variational wave method to the system with a QD embedded in a mesoscopic superconducting ring. The theoretical formulation of the variational wave approach are given in detail. We show  singlet-doublet transitions by tuning the QD energy level or the enclosed magnetic flux of the ring. We also calculate the Josephson currents in the ground state of this system, which show abrupt changes at the singlet-doublet transition points.

\section{Model Hamiltonian and the variational wave approach}

We consider a system consisted of a superconducting ring embedded with a quantum dot, which can be described by the following Hamiltonian
\bn
H&=&H_{\rm BCS}+H_{\rm DOT}+H_{\rm T}\;,\\
H_{\rm BCS}&=& -t\sum_{\sigma}\sum_{j=1}^{N-2}(c^\dagger_{j\sigma}c_{j+1\sigma}+\rm{h.c.})\nonumber\\
& -&\sum_{j=1}^{N-1} (\Delta e^{i\phi_s} c^\dagger_{j\uparrow}
c^\dagger_{j\downarrow}+\Delta e^{-i\phi_s}c_{j\downarrow}c_{j\uparrow})\;,\\
H_{\rm DOT}&=& \sum_\sigma \epsilon_d d^\dagger_\sigma d_\sigma +U n_{d\uparrow}n_{d\downarrow}\;,\\
H_{\rm T} &=& -t_L\sum_\sigma(d^\dagger_\sigma c_{1\sigma}+\rm{h.c.})\nonumber\\
& -&t_R\sum_\sigma (e^{i\phi_{ext}}c^\dagger_{N-1\sigma}d_\sigma+\rm{h.c.})\;.
\en
where $\Delta$ is the superconducting gap of the mesoscopic ring, $\phi_s$ is the phase of the superconductor order parameter. $\epsilon_d$
is the energy level of the dot, $U$ is the on-site Coulomb repulsion.  $t_L$ ($t_R$) denotes the hopping matrix element between the QD and the
left (right) neighboring site of the ring. The phase factor $\phi_{ext}$ is defined by $\phi_{ext}=2\pi\Phi/\Phi_0$, with $\Phi$ being the external magnetic
flux enclosed by the ring, and $\Phi_0$ being the flux quantum ($\Phi_0=hc/e$).

\begin{figure}[htp]
\includegraphics[width=0.8\columnwidth,height=2.5in,angle=0]{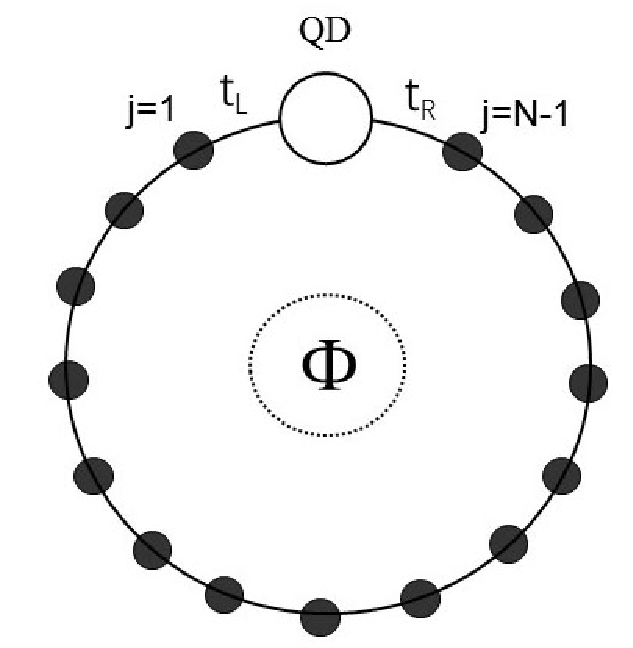}
\caption{ The schematic diagram of the superconducting ring-quantum dot system enclosing a magnetic flux $\Phi$.}
\end{figure}

By making a linear transformation
\bq
c_{j\sigma}=\sqrt{\frac{2}{N}}\sum_{m=1}^{N-1} c_{m\sigma} \sin(\frac{m\pi}{N}j), \hspace{0.5cm} (j=1,2,\cdots, N-1)\\
\eq
The Hamiltonian can be rewritten as
\bn
H &=&\sum_{\sigma}\sum_{m=1}^{N-1}\epsilon_m c^\dagger_{m\sigma}c_{m\sigma}
-\sum_{m=1}^{N-1} (\Delta e^{i\phi_s} c^\dagger_{m\uparrow}c^\dagger_{m\downarrow}\nonumber\\
&+&\Delta e^{-i\phi_s}c_{m\downarrow}c_{m\uparrow})+\sum_\sigma \epsilon_d d^\dagger_\sigma d_\sigma +U n_{d\uparrow}n_{d\downarrow}\nonumber\\
&+&\sum_\sigma\sum_{m=1}^{N-1}(t_m c^\dagger_{m\sigma}d_\sigma+t_m^* d^\dagger_\sigma c_{m\sigma})\;,
\en
with $\epsilon_m=-2t\cos(\frac{m\pi}{N})$, and
\bq
t_m=-\sqrt{\frac{2}{N}}[t_L+(-1)^{m+1}t_Re^{i\phi_{ext}}]\sin(\frac{m\pi}{N})\;.
\eq
The hybridization strength between the dot and the ring can be defined as: $\Gamma=\pi\sum_m|t_m|^2\delta(E_F-\epsilon_m)$.  By assuming the Fermi energy $E_F=0$, and in the large lattice number limit $N\rightarrow\infty$, we can obtain $\Gamma=(t_L^2+t_R^2)/t$.

Using the Bogoliubov transformation
\bn
c_{m\uparrow}&=&u_m \alpha_{m\uparrow}+v_m\alpha^\dagger_{m\downarrow}\;,\nonumber\\
c_{m\downarrow}&=&u_m \alpha_{m\downarrow}-v_m\alpha^\dagger_{m\uparrow}\;,
\en
where the factors: $u_m=\cos\theta_m $, $v_m=\sin \theta_m e^{i\phi_s}$, with
$\cos \theta_m=\frac{1}{\sqrt{2}}[1+{\epsilon_m/\sqrt{\epsilon_m^2+\Delta^2}}]^{1/2} $,
$\sin\theta_m=\frac{1}{\sqrt{2}}[1-{\epsilon_m/\sqrt{\epsilon_m^2+\Delta^2}}]^{1/2}$,
the BCS Hamiltonian can be expressed in terms of the quasiparticle operators
 \bq
 H_{\rm BCS}=\sum_{m,\sigma}E_m\alpha^\dagger_{m\sigma}\alpha_{m\sigma}\;,
\eq
where $E_m=\sqrt{\epsilon_m^2+\Delta^2} $ are the excitation energies of quasiparticles.  The electron tunneling part of the Hamiltonian can be rewritten as
\bn
 H_{\rm T}&=&\sum_{m}[t_m u_m(\alpha^\dagger_{m\uparrow}d_\uparrow+\alpha^\dagger_{m\downarrow}d_\downarrow)\nonumber\\
 &+&t_m v^*_m(\alpha_{m\downarrow}d_\uparrow-\alpha_{m\uparrow}d_\downarrow)]+\rm{h.c.}\;.
\en

The variational many-body states for the ground state of this system can be a singlet\cite{Rozhkov2000,Ilicin2025},
\begin{widetext}
\bq
| S\rangle =A\left [1+\sum_m\frac{1}{\sqrt{2}}B_m(\alpha^\dagger_{m\uparrow}d^\dagger_\downarrow
-\alpha^\dagger_{m\downarrow} d^\dagger_\uparrow)
+\sum_m\sum_{m'}(C_{m m'}\alpha^\dagger_{m\uparrow}\alpha^\dagger_{m'\downarrow}
+Q_{m m'}\alpha^\dagger_{m\uparrow}\alpha^\dagger_{m'\downarrow}d^\dagger_\uparrow d^\dagger_\downarrow)+D d^\dagger_\uparrow d^\dagger_\downarrow \right ]|0\rangle\;,
\eq
or a doublet with
\bn
| D\uparrow\rangle &=&\tilde A\left \{d^\dagger_\uparrow+\sum_m\tilde B_m\alpha^\dagger_{m\uparrow}
+\sum_m\sum_{m'}\left [\tilde C_{m m'}\alpha^\dagger_{m\uparrow}\alpha^\dagger_{m'\downarrow}d^\dagger_\uparrow
+\frac{1}{\sqrt{3}}\tilde D_{m m'}\alpha^\dagger_{m\uparrow}(\alpha^\dagger_{m'\uparrow}d^\dagger_\downarrow
-\alpha^\dagger_{m'\downarrow} d^\dagger_\uparrow)\right ]\right.\nonumber\\
&+&\left.\sum_m\tilde Q_m\alpha^\dagger_{m\uparrow} d^\dagger_\uparrow d^\dagger_\downarrow \right \}|0\rangle\;,
\en
and the energy degenerate state $| D\downarrow\rangle $.
\end{widetext}
It should be noted that the coefficients in the states: $C_{m m'}$, $Q_{m m'}$, $\tilde C_{m m'}$   are symmetric, e.g., $C_{m m'}=C_{m' m}$,  and the coefficient $\tilde D_{m m'}$ is antisymmetric:
$\tilde D_{m m'}=-\tilde D_{m' m}$. To obtain the lowest energy of the singlet, one can define a variational functional
\bq
F=\langle S|H|S\rangle-E_S\langle S|H|S\rangle\;,
\eq
with $E_S$ being the energy of the singlet state.
Then we set the functional derivatives with respect to the variational coefficients to be zero: $\frac {\delta F}{\delta C^*_{m m'}}=0$, $\frac {\delta F}{\delta Q^*_{m m'}}=0$, $\frac {\delta F}{\delta D^*}=0$, $\frac {\delta F}{\delta B^*_{m}}=0$. It gives rise to a set of self-consistent equations for the variational coefficients:
\bq
 C_{m m'}=\frac{t_m u_{m}B_{m'}
+t_{m'} u_{m'}B_{m}}{\sqrt{2}(E_S-E_m-E_{m'})}\;,
\eq
\bq
 Q_{m m'}=-\frac{t_m^* v_{m}B_{m'}
+t_{m'}^* v_{m'}B_{m}}{\sqrt{2}(E_S-2\epsilon_d-U-E_{m}-E_{m'})}\;,
\eq
\bq
D=\frac{1}{E_S-2\epsilon_d-U}\sum_{m}\sqrt{2}t_m^* u_{m}B_{m}\;,
\eq
and
\bn
B_{m}&=&\frac {\sqrt{2}}{E_S-\epsilon_d-E_{m}}[ t^*_{m}v_{m}+t_{m} u_{m}D\nonumber\\
&+& \sum_{m'} (t^*_{m'}u_{m'}C_{m m'}
-t_{m'}v_{m'}^*Q_{m m'} )  ]\;.
\en
By substituting Eqs.(14)-(16) into Eq.(17), we obtain the linear equations satisfied by the coefficients $B_{m}$,
\begin{widetext}
\bn
& &\left [E_S-E_{m}-\epsilon_d-\sum_{m'}\left (\frac{|t_{m'}|^2 u_{m'}^2}{E_S-E_{m}-E_{m'}}
+\frac{|t_{m'}|^2 |v_{m'}|^2}{E_S-2\epsilon_d-U-E_{m}-E_{m'}}
\right ) \right ]B_{m}
=\sqrt{2}t^*_{m}v_{m}\nonumber\\
& &\hspace {2cm} +\sum_{m'}\left (\frac{t_m u_{m}t^*_{m'}u_{m'}}{E_S-E_{m}-E_{m'}}
+\frac{t_m^* v_{m}t_{m'}v_{m'}^*}{E_S-2\epsilon_d-U-E_{m}-E_{m'}}+ \frac{2 t_m u_m t^*_{m'}u_{m'}}{E_S-2\epsilon_d-U} \right  )B_{m'}\;.
\en
\end{widetext}
The energy of the singlet state will be given as
\bq
E_S=\sum_{m}\sqrt{2} t_m v_{m}^*B_{m}\;.
\eq
Eqs. (18) and (19) form the set of self-consistent equations for the singlet state, and will be solved numerically.

In a similar procedure, we can also obtain a set of self-consistent equations for the variational coefficients: $\tilde B_{m}$, $\tilde C_{m m'}$,$\tilde D_{m m'}$, $\tilde Q_{m}$ and the energy $E_D$ of the doublet state (The detailed expressions are given in Appendix A).

\section{The singlet-doublet transition and Josephson currents}

  We solve the self-consistent equations for the singlet and the doublet numerically, and determine all the variational coefficients and the ground state energies $E_S$ and $E_D$.   In our calculations, we take the total lattice number of the ring-dot system: $N=1000$, The on-site Coulomb repulsion  $U=4.0\Gamma $. The other model parameters: $t=1.0, t_L=t_R=0.1$, which leads to the hybridization strength $\Gamma=0.02$.

\begin{figure}[htp]
\includegraphics[width=\columnwidth,height=4.0in,angle=0]{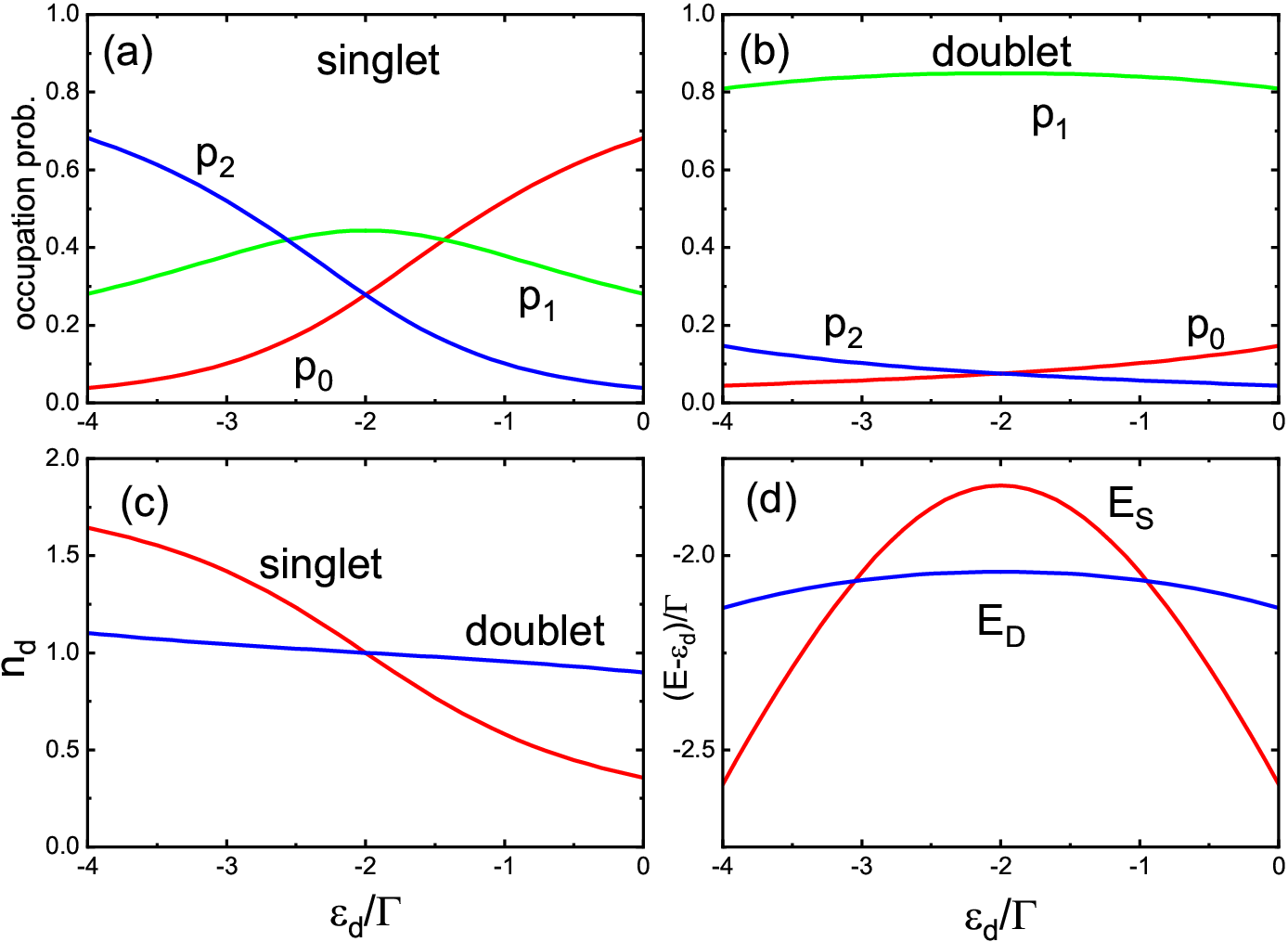}
\caption{ The characteristic properties of the singlet and the doublet vs. the energy level of the dot. (a) the probabilities of the dot level in the empty, single occupation,  double occupation states for the singlet state. (b) That of the doublet state. (c) The total occupation number of electrons on the dot level. (d) The lowest energy of the singlet and the doublet vs. the dot level. In our calculations (except in Figure 3),  we always take the superconducting gap $\Delta=2.0\Gamma$ and the order parameter phase factor $\phi_s=0$.}
\end{figure}
At first, we consider the systems without the enclosed magnetic flux ($\phi_{ext}=0$). To ensure the numerical results of calculation to be reliable, we show the characteristic properties of the singlet and the doublet
in Figure 2.  The probabilities of empty, singly occupied(with spin up and down) and doubly occupied electronic states at the QD: $p_0, p_1, p_2$, respectively,  are plotted for the singlet and the doublet in Figure 2(a) and(b).  One can see that the changes of the probabilities $p_0, p_1, p_2$  versus the dot level $\epsilon_d $ are in agreement with the physical meaning  of the the singlet and the doublet, e.g.,  the probability $p_1$ of singly occupied state dominates over $p_0$ and $p_2$ in the doublet,  all of the occupation probabilities exhibit the particle-hole symmetry with respective to the dot level point $\epsilon_d=-U/2$. The total occupation number $n_d$ of the QD versus the dot level $\epsilon_d$ is shown in Figure 2(c). We see that the dot occupation number $n_d$ changes obviously with the tuning of the dot level $\epsilon_d$ in the singlet, but it is almost fixed at $n_d\approx 1$  in the doublet. In Figure 2(d) the energies of the singlet and the doublet  versus the dot level $\epsilon_d$ are plotted. We find that in the particle-hole symmetric region ($\epsilon_d \approx -U/2$),  the energy of the doublet $E_D$ is lower than that of the singlet $E_S$, the ground state of this system is a doublet. But in the mixed valence region (with $\epsilon_d\approx 0$ or $\epsilon_d\approx -U  $),  the energy of the singlet $E_S$ is lower than that of the doublet $E_D$. Therefore, the system should exhibit singlet-doublet phase transitions by tuning the dot level $\epsilon_d$.

\begin{figure}[htp]
\includegraphics[width=\columnwidth,height=3.0in,angle=0]{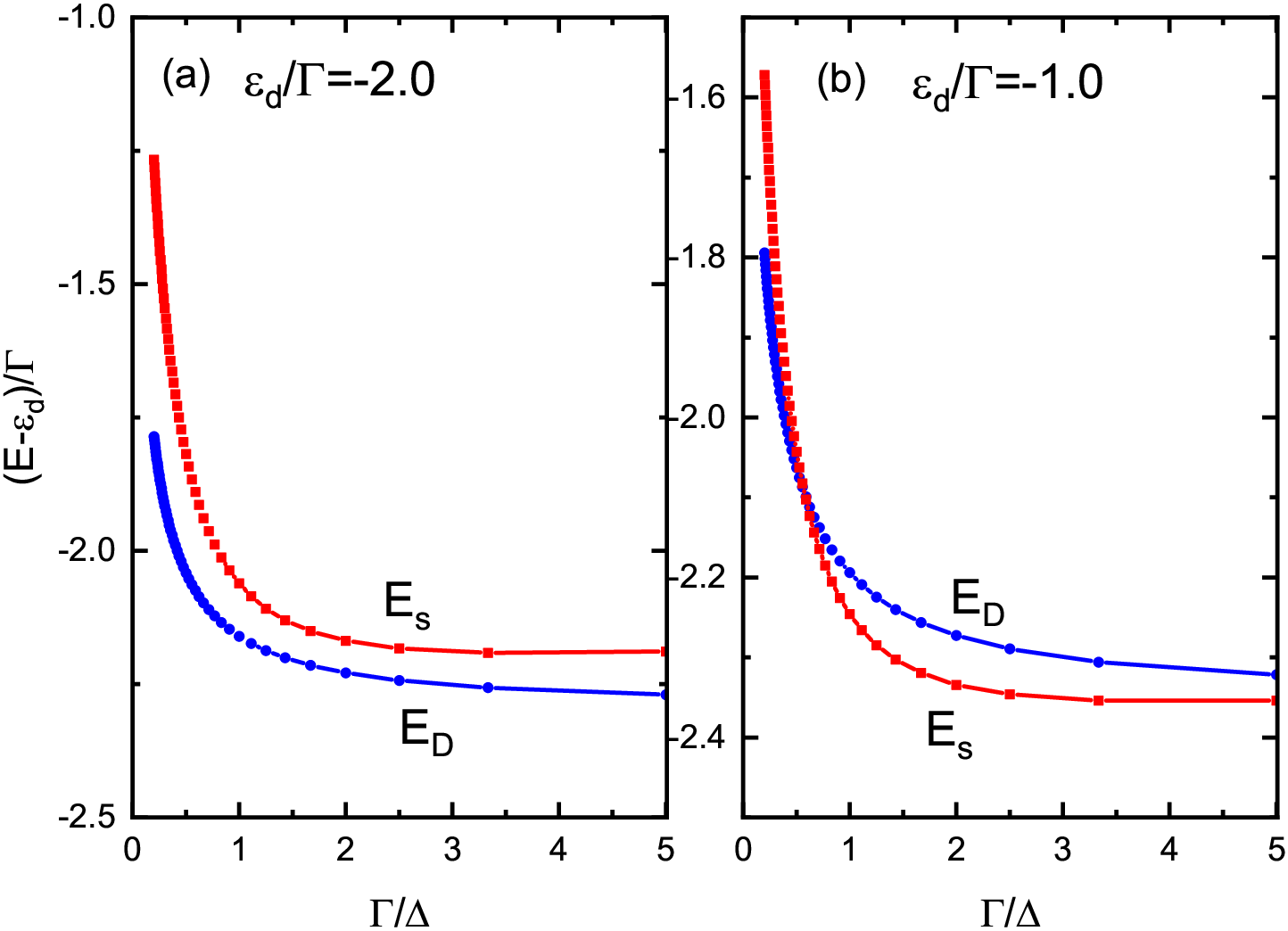}
\caption{The energies $E_S$ and $E_D$ of the singlet and the doublet, respectively,  vs. the superconducting gap $\Delta$. }
\end{figure}

 In Figure 3,  we study the evolutions of the energies $E_S$ and $E_D$ with the change of the magnitude of the superconducting gap $\Delta$ by fixing the hybridization strength $\Gamma$.   As indicated in Figure 3(a),  the energy
of the doublet is always lower than that of the singlet in the particle-hole symmetric region even with $\Delta\ll \Gamma$. But in the mixed valence region as shown in Figure 3(b),  there is a doublet to singlet transition of the ground state with decreasing of the magnitude of $\Delta$, and it may be regarded as a phase transition from the ABS to the YSR singlet.

Next we study the dependence of the energies  $E_S$ and $E_D$ on the enclosed magnetic flux $\Phi$. Energy versus the phase $\phi_{ext}$ induced by external magnetic flux is plotted in Figure 4 at different values of the dot level $\epsilon_d$. One can see that when $\epsilon_d/\Gamma=-1.0$, the doublet is the ground state for all $\phi_{ext}$, while for $\epsilon_d/\Gamma=-0.4$, the ground state is always the singlet state.
The ground state energy function $E(\phi_{ext})$ exhibits a periodic structure with the period being $\pi$ instead of $2\pi$, which can be attributed to the fact that the carriers in the superconducting ring are Cooper's pairs with charge $2e$. This $\pi$ periodic structure is different from that for the system of a normal metallic ring embedded with a QD, in which the ground state energy and the persistent current have a period $2\pi$ with respect to the phase $\phi_{ext}$.\cite{Ding2003a} , and is also different from the ground state energy dependence of QD Josephson junctions on the phase difference $\Delta \phi_s$ between two superconducting electrodes.\cite{Rozhkov2000} When $\epsilon_d/\Gamma=-0.9$, the energy curves of $E_S$ and $E_D$ cross each other, and the ground state energy has kinks as a function of the phase $\phi_{ext}$. Therefore, $E(\phi_{ext})$ has two local minima at $\phi_{ext}=0$ and $\phi_{ext}=\pi/2$,  with $\phi_{ext}=\pi/2$ being the global minimum. When $\epsilon_d/\Gamma=-0.7$, the ground state energy $E(\phi_{ext})$ also kink structures and  two local minima at $\phi_{ext}=0$ and $\phi_{ext}=\pi/2$,  but with $\phi_{ext}=0$ being the global minimum.

\begin{figure}[htp]
\includegraphics[width=\columnwidth,height=4.0in,angle=0]{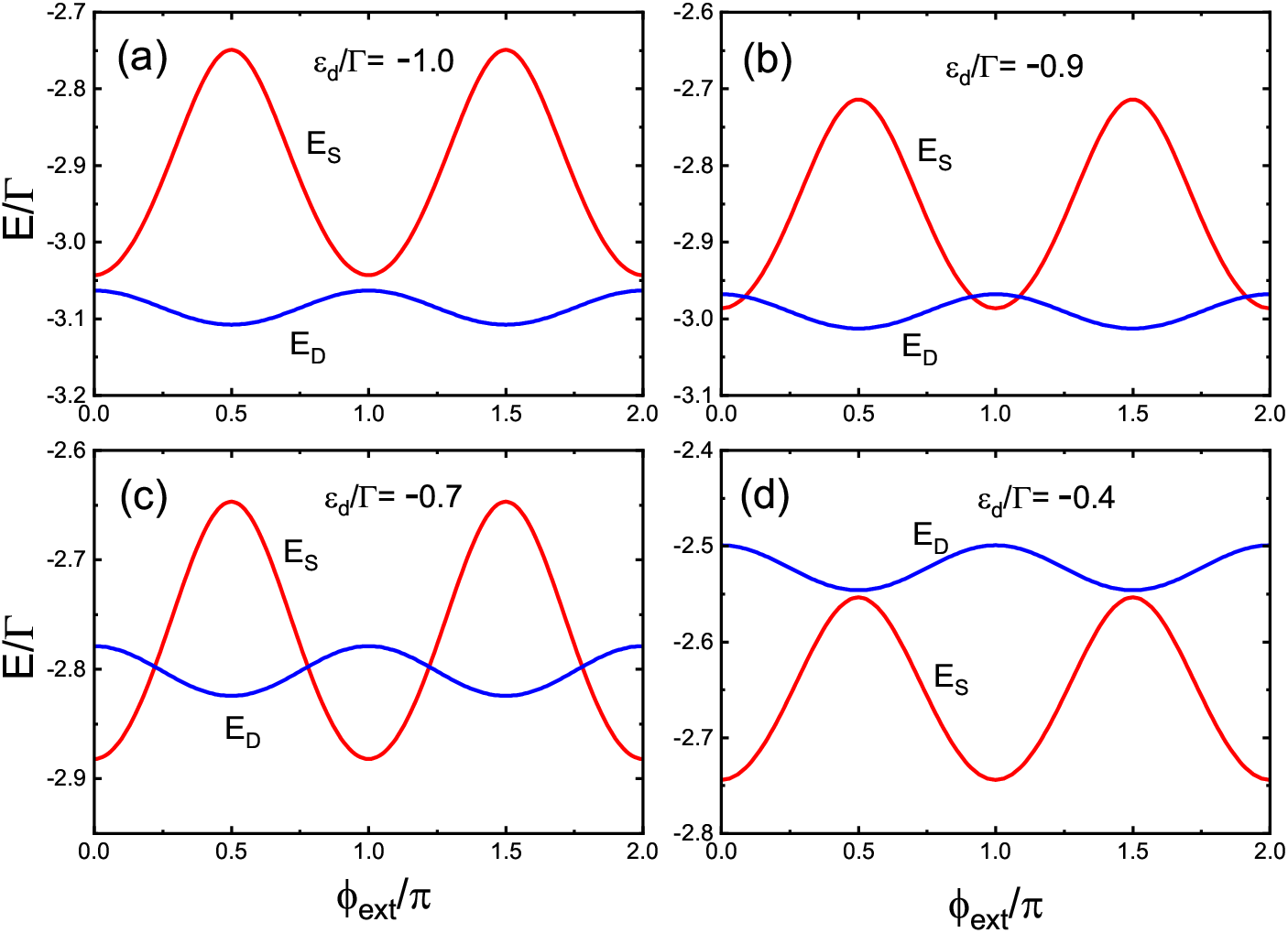}
\caption{The energies $E_S$ and $E_D$ of the singlet and the doublet, respectively,  vs. the enclosed magnetic flux $\Phi$ of the ring (with $\phi_{ext}=2\pi \Phi/\Phi_0$).  }
\end{figure}

\begin{figure}[htp]
\includegraphics[width=\columnwidth,height=4.0in,angle=0]{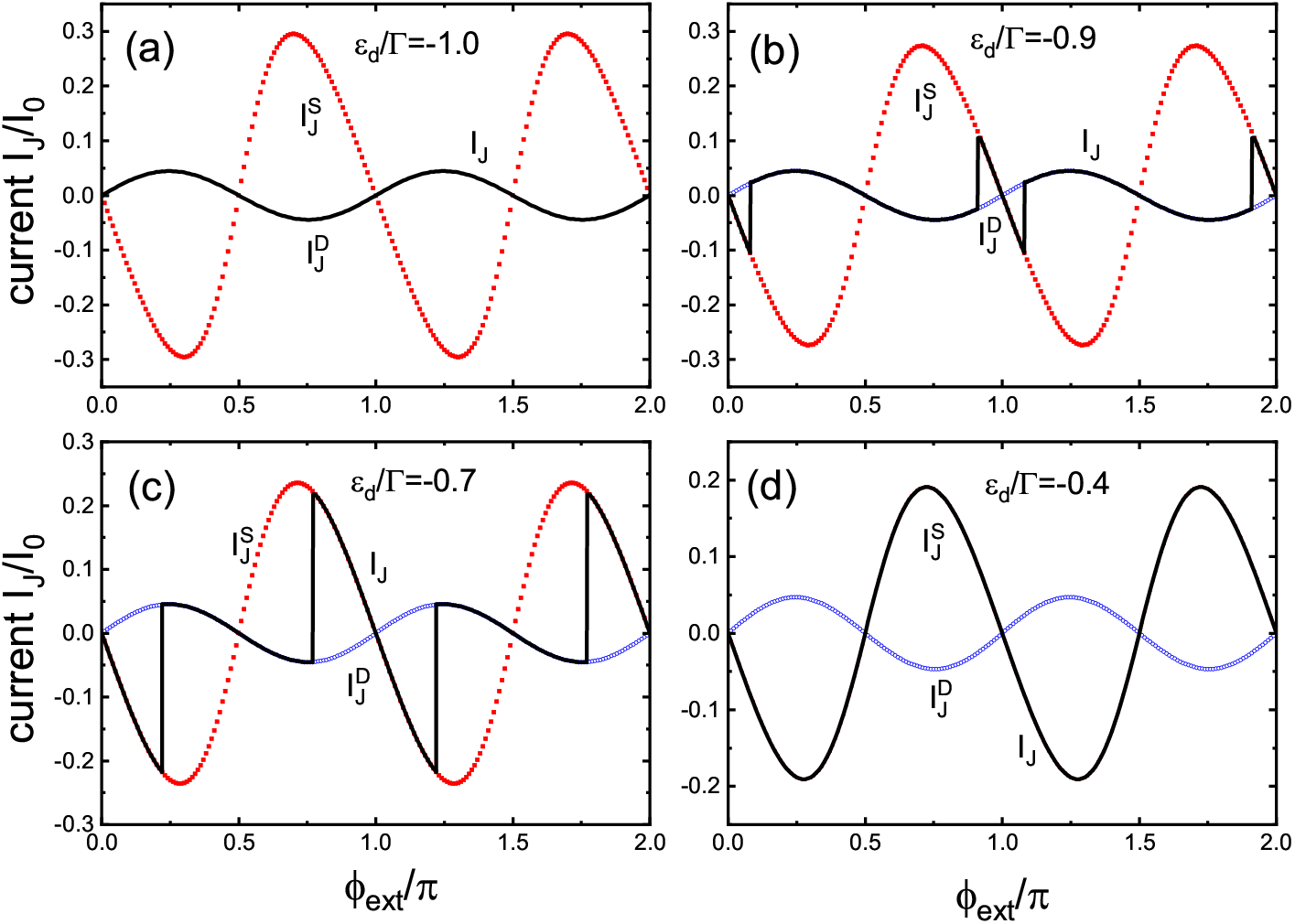}
\caption{The Josephson currents  vs. the enclosed magnetic flux $\Phi$ of the ring (with $\phi_{ext}=2\pi \Phi/\Phi_0$), where $I_0=e\Gamma/\hbar$. }
\end{figure}

The Josephson current $I_J$ of the superconducting ring at zero temperature is given by the expectation values of the current operator in the ground state as follows
\bn
I_{J}&=&-\frac{e}{\hbar}\frac{\partial E(\phi_{ext})}{\partial \phi_{ext}}
\nonumber\\
&=&\frac{ie}{\hbar}\sum_{\sigma}\langle \psi_G|t_R e^{i\phi_{ext}} c^\dagger_{N-1\sigma}d_\sigma-{\rm h.c.}|\psi_G\rangle\;,
\en
with $|\psi_G\rangle=|S\rangle$ or $|D\uparrow\rangle$. The current $I_J$ can be calculated by rewriting the current operator in terms of the quasiparticle operators (the details
are given in Appendix B). In Figure 5 we plot the Josephson currents versus the external magnetic flux $\Phi$ at different values of $\epsilon_d$.  Making comparison between Figure 5(a) and (d), we see that the magnitude of the Josephson current $I_J$ becomes significantly larger when the ground state of this system is a singlet than a doublet, and also the direction of the current is reversed.  As shown in Figure 5(b) and (c), the singlet-doublet transitions of the ground state are also manifested as abrupt jumps of the Josephson current at some particular values of the phase $\phi_{ext}$.

\section{Conclusion}
In conclusion, we have studied the singlet-doublet transition and Josephson currents in the system with a superconducting ring embedded with a QD by a finite-$U$ variational wave function method. The benefits of the wave function approach are the easy implement of this method and its clear physical picture of the variational wave function. We show that this method does capture the essential features of the singlet-transition and Joesphson currents in the ground state. It should be noted that our calculation is limited to a superconductiong ring with finite length, in which the energy level discreteness plays an important role, therefore the Kondo regime  might not be well described. It is also noted that in the wave function ansatz applied in this paper only the two quasiparticle states in the electrodes are take into account, therefore it might be appropriate only to the systems with weak or intermediate tunnelling coupling between the QD and the electrodes. For the case with strong hybridization strength between the QD and the electrodes, some nonperturbative methods, e.g., the NRG methods and the quantum Monte carlo method,  have to be utilized.  For the consideration of the Kondo effect in the QD josephson juntions, it will be crucial to properly take into account the continuum band structure of the superconducting electrodes\cite{Ilicin2025}. In further work,  we may also expect this wave function method can be generalized to the systems with the multi-orbital electrodes.

\begin{acknowledgments}
{F. Ye thanks the support by the National Natural Science Foundation of China (Grant No. 11774143). }
\end{acknowledgments}

\newpage
\appendix
\begin{appendix}
\section{}
We can define a functional for the doublet as
\bq
F=\langle D\uparrow|H|D\uparrow \rangle-E_D\langle D\uparrow|H|D\uparrow\rangle\;,
\eq
where $E_D$ is the energy of the doublet. Then the functional derivative equations: $\frac {\delta F}{\delta \tilde B^*_{m}}=0$,
$\frac {\delta F}{\delta \tilde C^*_{m m'}}=0$, $\frac {\delta F}{\delta \tilde D^*_{m m'}}=0$, $\frac {\delta F}{\delta \tilde Q^*_{m}}=0$, will lead to a set of self-consistent equations as follows
\bn
 \tilde C_{m m'}&=&-\frac{1}{2}\left [\frac{t_{m}^* v_{m}\tilde B_{m'}+t_{m'}^* v_{m'}\tilde B_{m}}{E_D-\epsilon_d-E_{m}-E_{m'}}\right.\nonumber\\
 & &+\left. \frac{t_{m} u_{m}\tilde Q_{m'}+t_{m'} u_{m'}\tilde Q_{m}}{E_D-\epsilon_d-E_{m}-E_{m'}}\right ] \;,
\en
\bn
 \tilde D_{m m'}&=&-\frac{\sqrt{3}}{2}\left [\frac{t_{m}^* v_{m}\tilde B_{m'}-t_{m'}^* v_{m'}\tilde B_{m}}{E_D-\epsilon_d-E_{m}-E_{m'}}\right.\nonumber\\
& &+\left. \frac{t_{m} u_{m}\tilde Q_{m'}-t_{m'} u_{m'}\tilde Q_{m}}{E_D-\epsilon_d-E_{m}-E_{m'}}\right ]\;,
\en
\bq
\tilde B_{m}=\frac { t_{m} u_{m}
-\sum_{m'}t_{m'}v_{m'}^*(\sqrt{3}\tilde D_{m' m}+\tilde C_{m' m}) } {E_D-E_{m}} \;.
\eq
\bq
\tilde Q_{m}=\frac { -t_{m}^* v_{m}
-\sum_{m'} t^*_{m'}u_{m'}(\sqrt{3}\tilde D_{m' m}+\tilde C_{m' m}) } {E_D-2\epsilon_d-U-E_{m}} \;.
\eq
By substituting Eqs. A(2), A(3) into Eqs. A(4), A(5), we can obtain the linear Eqs. satisfied by the coefficients $\tilde B_{m}$ and
 $\tilde Q_{m}$,
\begin{widetext}
\bn
\left (E_D-E_{m}-\sum_{m'}\frac{2|t_{m'}|^2 |v_{m'}|^2}{E_D-\epsilon_d-E_{m}-E_{m'}}
 \right )\tilde B_{m}=t_{m}u_{m}
-\sum_{m'}\frac{t_m^* v_{m}t_{m'}v_{m'}^*\tilde B_{m'}}
{E_D-\epsilon_d-E_{m}-E_{m'}}\nonumber\\
 +\sum_{m'}\frac{2t_{m'} u_{m'}t_{m'}v_{m'}^*\tilde Q_{m}-t_m u_{m}t_{m'}v_{m'}^*\tilde Q_{m'}}{E_D-\epsilon_d-E_{m}-E_{m'}}\hspace{2cm}\;,
\en
\bn
\left (E_D-2\epsilon_d-U-E_{m}-\sum_{m'}\frac{2|t_{m'}|^2 u_{m'}^2}{E_D-\epsilon_d-E_{m}-E_{m'}} \right )\tilde Q_{m}=-t_{m}^*v_{m}
-\sum_{m'}\frac{t_m u_{m}t_{m'}^*u_{m'}\tilde Q_{m'}}
{E_D-\epsilon_d-E_{m}-E_{m'}}\nonumber\\
 +\sum_{m'}\frac{2t_{m'}^*v_{m'}t_{m'}^*u_{m'}\tilde B_{m}
 -t_m^* v_{m}t_{m'}^*u_{m'}\tilde B_{m'}}{E_D-\epsilon_d-E_{m}-E_{m'}}\hspace{3cm}\;.
\en
\end{widetext}
The energy of the doublet will be given as
\bq
E_D=\epsilon_d+\sum_{m}(t_m^* u_{m}\tilde B_{m}-t_m v_{m}^*\tilde Q_{m})\;.
\eq

\section{}
The current in the singlet state is given by
\bq
I_{J}^{S}=\frac{ie}{\hbar}\sum_{\sigma}\langle S|t_R e^{i\phi_{ext}} c^\dagger_{N-1\sigma}d_\sigma-{\rm h.c.}|S\rangle\;.
\eq
It can be rewritten in terms of the quasiparticle operators, and be separated to two parts,
\bq
I_J^{S}=I_{S}^{qp}+I_{S}^{pair}\;,
\eq
\bn
I_{S}^{qp}&=&\frac{ie}{\hbar}\sqrt{\frac{2}{N}}\sum_{m}(-1)^{m+1}\sin(\frac{m\pi}{N})\left [ t_R e^{i\phi_{ext}} u_{m} \right.\nonumber\\
&& \left.\cdot \langle S |(\alpha^\dagger_{m \uparrow}d_\uparrow+\alpha^\dagger_{m \downarrow}d_\downarrow)|S\rangle-{\rm c.c}\right ]\;,
\en
\bn
I_{S}^{pair}&=&\frac{ie}{\hbar}\sqrt{\frac{2}{N}}\sum_{m}(-1)^{m+1}\sin(\frac{m\pi}{N})\left [ t_R e^{i\phi_{ext}} v_{m}^* \right.\nonumber\\
&& \left.\cdot \langle S |(\alpha_{m \downarrow}d_\uparrow-\alpha_{m \uparrow}d_\downarrow)|S\rangle-{\rm c.c}\right ]\;.
\en
They can be explicitly expressed in terms of the variational coefficients in the singlet state,
\begin{widetext}
\bq
I_{S}^{qp}=\frac{ie}{\hbar}|A|^2 \left [\sqrt{\frac{2}{N}}\sum_{m}(-1)^{m+1}\sin(\frac{m\pi}{N})t_R e^{i\phi_{ext}}u_{m} \left (\sqrt{2} B_{m}^*D+ \sum_{m'} \sqrt{2}C_{m m'}^*B_{m'}
\right ) -{\rm c.c.} \right ]\;,
\eq
\bq
I_{S}^{pair}=\frac{ie}{\hbar}|A|^2 \left [\sqrt{\frac{2}{N}}\sum_{m}(-1)^{m+1}\sin(\frac{m\pi}{N})t_R e^{i\phi_{ext}}v_{m}^* \left (\sqrt{2} B_{m}-\sum_{m'} \sqrt{2}Q_{m m'}B_{m'}^*
\right ) -{\rm c.c.} \right ]\;.
\eq
We see that the Josephson current $I_{J}^{S}$ has contributions from the quasiparticle tunneling current $I_S^{qp}$ and also the paired tunneling current $I_S^{pair}$. In a similar way, one can take the expectation value of  current operator in the doublet state, and obtain the Josephson current of a doublet state,
\bq
I_J^{D}=I_{D}^{qp}+I_{D}^{pair}\;,
\eq
\bq
I_{D}^{qp}=\frac{ie}{\hbar}|\tilde A|^2 \left [\sqrt{\frac{2}{N}}\sum_{m}(-1)^{m+1}\sin(\frac{m\pi}{N})t_R e^{i\phi_{ext}}u_{m} \left (\tilde B_{m}^*- \sum_{m'} (\sqrt{3}\tilde D_{m m'}^*+\tilde C_{m m'}^*)\tilde Q_{m'}
\right ) -{\rm c.c.} \right ]\;,
\eq
\bq
I_{D}^{pair}=\frac{ie}{\hbar}|\tilde A|^2 \left [\sqrt{\frac{2}{N}}\sum_{m}(-1)^{m+1}\sin(\frac{m\pi}{N})t_R e^{i\phi_{ext}}v_{m}^* \left (-\tilde Q_{m}- \sum_{m'} (\sqrt{3}\tilde D_{m m'}+\tilde C_{m m'})\tilde B_{m'}^*
\right ) -{\rm c.c.} \right ]\;.
\eq
\end{widetext}

\end{appendix}

\bibliography{RefsRingDot}

\end{document}